\documentclass[acmsmall,screen]{acmart}
\AtBeginDocument{%
  \providecommand\BibTeX{{%
    Bib\TeX}}}

\setcopyright{acmlicensed}
\copyrightyear{2018}
\acmYear{2018}
\acmDOI{XXXXXXX.XXXXXXX}

\acmJournal{JACM}
\acmVolume{37}
\acmNumber{4}
\acmArticle{111}
\acmMonth{8}

\usepackage{makecell,fancyvrb}

\let\digamma\relax

\usepackage{amsmath,amssymb,amsfonts}
\usepackage{algorithmic}
\usepackage{graphicx}
\usepackage{textcomp}
\usepackage{xcolor}
\usepackage{makecell}
\usepackage{multirow}
\usepackage[normalem]{ulem}
\def\BibTeX{{\rm B\kern-.05em{\sc i\kern-.025em b}\kern-.08em
    T\kern-.1667em\lower.7ex\hbox{E}\kern-.125emX}}
\newcommand{\MCot}{\textit{MelcotCR}}

\usepackage{caption}
\begin{document}

\title{Fine-Tuning LLMs to Analyze Multiple Dimensions of Code Review: A Maximum Entropy Regulated Long Chain-of-Thought Approach}


\author{Yongda Yu}
\affiliation{%
  \institution{Nanjing University}
  \city{Nanjing}
  \country{China}}
\email{yuyongda@smail.nju.edu.cn}

\author{Guohao Shi}
\affiliation{%
  \institution{Nanjing University}
  \city{Nanjing}
  \country{China}}
\email{335933870@qq.com}

\author{Xianwei Wu}
\affiliation{%
  \institution{Nanjing University}
  \city{Nanjing}
  \country{China}}
\email{652024320006@smail.nju.edu.cn}

\author{Haochuan He}
\affiliation{%
  \institution{Nanjing University}
  \city{Nanjing}
  \country{China}}
\email{1516998446@qq.com}

\author{XueMing Gu}
\affiliation{%
  \institution{University of Waterloo}
  \city{Avenue}
  \country{Canada}}
\email{x7gu@uwaterloo.ca}

\author{Qianqian Zhao}
\affiliation{%
  \institution{Northeastern University}
  \city{Shenyang}
  \country{China}}
\email{zhaoqianqian2025@163.com}

\author{Kui Liu}
\affiliation{%
 \institution{Huawei Technologies Co., Ltd.}
 \city{Xi'an}
 \country{China}}
\email{kui.liu@huawei.com}

\author{Qiushi Wang}
\affiliation{%
 \institution{Huawei Technologies Co., Ltd.}
 \city{Xi'an}
 \country{China}}
\email{wangqiushi6@huawei.com}

\author{Zhao Tian}
\affiliation{%
 \institution{Huawei Technologies Co., Ltd.}
 \city{Xi'an}
 \country{China}}
\email{tianzhao@huawei.com}

\author{Haifeng Shen}
\affiliation{%
 \institution{Southern Cross University}
 \city{Gold Coast}
 \country{Australia}}
\email{haifeng.shen@scu.edu.au}

\author{Guoping Rong}
\authornote{Guoping Rong is the corresponding author.}
\email{ronggp@nju.edu.cn}
\affiliation{%
  \institution{Nanjing University}
  \city{Nanjing}
  \country{China}
}

\renewcommand{\shortauthors}{Yu et al.}

\begin{abstract}
Large Language Models (LLMs) have shown great potential in supporting automated code review due to their impressive capabilities in context understanding and reasoning. However, these capabilities are still limited compared to human-level cognition because they are heavily influenced by the training data. Recent research has demonstrated significantly improved performance through fine-tuning LLMs with code review data. However, compared to human reviewers who often simultaneously analyze multiple dimensions of code review to better identify issues, the full potential of these methods is hampered by the limited or vague information used to fine-tune the models. This paper contributes \MCot, a chain-of-thought (COT) fine-tuning approach that trains LLMs with an impressive reasoning ability to analyze multiple dimensions of code review by harnessing long COT techniques to provide rich structured information. To address context loss and reasoning logic loss issues that frequently occur when LLMs process long COT prompts, we propose a solution that combines the Maximum Entropy (ME) modeling principle with pre-defined reasoning pathways in \MCot\ to enable more effective utilization of in-context knowledge within long COT prompts while strengthening the logical tightness of the reasoning process. Empirical evaluations on our curated \MCot\ dataset and the public CodeReviewer dataset reveal that a low-parameter base model, such as 14B Qwen2.5, fine-tuned with \MCot\ can surpass state-of-the-art methods in terms of the accuracy of detecting and describing code issues, with its performance remarkably on par with that of the 671B DeepSeek-R1 model.
\end{abstract}


\renewcommand{\shorttitle}{\MCot}



\maketitle

\section{Introduction}
As a cornerstone practice in software quality assurance, code review plays an irreplaceable role in modern software development~\cite{gousios2014exploratory}. 
Traditional manual review processes, which heavily rely on line-by-line scrutiny of code by developers, are not only time-consuming and labor-intensive but also heavily dependent on individual reviewers' expertise, a limitation that has driven the innovation and advancement of Automated Code Review (ACR) technologies~\cite{li2022automating, tufano2021towards, lu2023llama, yu2024fine, gupta2018intelligent}. Recent breakthroughs in AI-driven code analysis based on Large Language Models (LLMs) have naturally attracted significant research attention to ACR, resulting in multiple LLM-based ACR systems~\cite{lu2023llama, yu2024fine}. 

As an LLM's capability in context understanding and reasoning is highly influenced by its training data, a common strategy is to fine-tune base LLMs with code-review-specific data so that they can perform better in ACR tasks. In particular, Liu et al.~\cite{lu2023llama} fine-tuned Llama to create Llama-Reviewer, which is able to generate review comments through the direct learning of a raw code review dataset from a previous study~\cite{li2022codereviewer} to enhance the LLM's commenting capability. It should be noted that the information used in fine-tuning is only a direct code-to-comment mapping that lacks analytical depth. In contrast, Yu et al. developed Carllm~\cite{yu2024fine} that used chain-of-thought (COT) to fine-tune the base LLMs by decomposing the review process into three progressive stages: defect localization, review comment generation, and solution proposal. The COT approach provides richer information for fine-tuning, resulting in improved ACR performance compared to Llama-Reviewer.

However, compared to human reviewers who often simultaneously consider multiple dimensions of code review, e.g., code intent, boundary conditions, and invocation relationships, when analyzing code issues~\cite{bosu2013impact,kononenko2016code}, the full potential of LLMs for ACR is hampered by the limited or vague information used to fine-tune them. Specifically,  Llama-Reviewer~\cite{lu2023llama} merely instructs the LLM to perform ACR via vague prompts, while in Carllm~\cite{yu2024fine}, each reasoning step is constrained in such a way that it only requires the LLM to identify issues without providing much guiding information. Some studies have revealed a trend that employing COT to prompt LLMs can significantly enhance their performance in specific tasks~\cite{wei2022chain, feng2023towards, zhang2022automatic}, while there exists a positive correlation between COT length and model performance gain~\cite{chen2025towards, xia2024beyond}.

In this paper, we contribute \MCot, a fine-tuning approach that trains LLMs with an impressive structured reasoning ability to analyze multiple dimensions of code review by harnessing long COT techniques~\cite{wu2025more, chen2025towards, yeo2025demystifying}. It systematically decomposes a code review task into multiple fine-grained sub-tasks including code functionality summarization, core logic analysis, change impact analysis, and inspection of multiple concrete issues, thereby stimulating an LLM to generate coherent reasoning chains. However, a known challenge with long COT is its inherent context loss and reasoning logic loss issues, as the
length of COT increases
~\cite{chen2025towards, xia2024beyond}. 

% \paragraph{Challenge}\R{What is the top chellenge in this paper?}

To address this challenge, we propose a solution that combines the Maximum Entropy (ME) modeling principle with pre-defined reasoning pathways in \MCot\ to enable more effective utilization of in-context knowledge within long COT prompts while strengthening the logical tightness of the reasoning process. This solution expands each ground truth answer into multiple semantically equivalent but syntactically distinct expressions, enabling the model to learn bias-invariant essential knowledge.

The main contributions of this work are summarized as follows:
\begin{itemize}
    \item We propose \MCot, a novel fine-tuning approach that trains LLMs with the ability to analyze multiple dimensions in code review by combining the maximum entropy principle with long COT techniques to cultivate diversified problem-solving capabilities.
    \item We conduct empirical evaluations to demonstrate that a 14B low-parameter base model fine-tuned with \MCot\ can outperform state-of-the-art methods, with its ACR performance remarkably on par with that of Deepseek-R1 671B.
    \item We perform experiments to investigate the impacts of fine-tuning methods and reasoning steps on \MCot's ACR performance. The findings not only validate the proposed approach but also suggest future research directions.
\end{itemize}

The rest of the paper is organized as follows. Section~\ref{sec:relat} introduces some related work. Section~\ref{sec:meth} describes the research methodology, followed by the evaluation process and results in Section~\ref{sec:eval}. Section~\ref{sec:disc} discusses the implications and validity risks. Section~\ref{sec:conc} concludes the paper with a summary of contributions and future work.

\section{Related work}
\label{sec:relat}
In this section, we review the related work that forms the foundation of this work, including automated code review, chain of thought, and maximum entropy.

\subsection{Automated Code Review (ACR)}
ACR is an essential aspect of improving software development efficiency, which aims to reduce the manual effort and time required for code assessment. The main focus of an ACR system is to detect potential code defects and suggest or generate relevant review comments. It typically comprises two components: defect detection and review comment recommendation/generation. 

Defect detection aims to identify potential issues within code snippets under review. For instance, CodeT5~\cite{wang2021codet5} adopts a unified framework that supports both code understanding and generation tasks, thus facilitating multi-task learning. CodeBert~\cite{feng2020codebert} is a bimodal pre-training model tailored for programming languages and natural language, excelling in tasks such as natural language-based code search and code documentation generation. DACE~\cite{shi2019automatic} employs CNN and LSTM techniques to extract Diff features from the code, enabling the prediction of code quality in Diff patches. LogiCode~\cite{zhang2024logicode} leverages LLMs to detect logical anomalies, automatically generating Python code to identify issues such as incorrect component quantities or missing elements.

Review comment recommendation/generation produces review comments through retrieval or generation methods. For example, LLaMA-Reviewer~\cite{lu2023llama} uses low-parameter fine-tuning techniques to enhance LLaMA, leading to impressive results in generating review comments. CodeReviewer~\cite{li2022codereviewer} achieves notable success in detecting code defects, generating review comments, and performing code repair tasks by developing pre-training tasks specifically designed for code review in an end-to-end manner. Notably, both studies utilize the same dataset~\cite{li2022codereviewer} for training and validating their models, assuming that the existence of review comments represents the ground truth, without evaluating whether these comments are genuinely related to the code fixes. DCR~\cite{gupta2018intelligent} learns the similarity between code commit `diffs' and review comments to retrieve comments relevant to specific code commits. CommentFinder ~\cite{hong2022commentfinder} employs deep learning techniques to retrieve pertinent code review comments, thus minimizing the time reviewers spend crafting these comments. The BitsAI-CR~\cite{ning2024defining} framework enhances ACR through a two-stage approach that combines a rule-based initial issue detection with a model called Reviewfilter for verification.  This system, implemented with a taxonomy of review rules,  achieves a precision rate of 75.0\% in review comment generation.
% while maintaining an Outdated Rate of 26.7\% for the Go language at ByteDance.

\subsection{Chain of Thought (COT)}
COT is commonly applied in LLMs to assist in better problem-solving or decision-making~\cite{wei2022chain}. This process helps an LLM generate more accurate, context-aware, and logical output, particularly in complex tasks like code review, debugging, or code generation~\cite{10.1145/3690635, yu2024fine, nong2024chain}. COT requires breaking down a problem into smaller and manageable components and guiding the model to reason through each part step by step. For example, when analyzing code to identify potential issues, the model will first pinpoint the location of a potential problem, then describe the specific issue at that location, and finally suggest a repair solution~\cite{yu2024fine}. This sequential reasoning process helps improve the overall accuracy of the model's output and ensures that it understands the underlying issue, not just generating a quick response. 
In addition, Yu Nong et al.~\cite{nong2024chain} investigated the use of LLMs and the COT prompting to address security vulnerabilities in software, which achieved impressive results. 

\begin{figure*}[!t]
    \centering
    \includegraphics[width=0.95\linewidth]{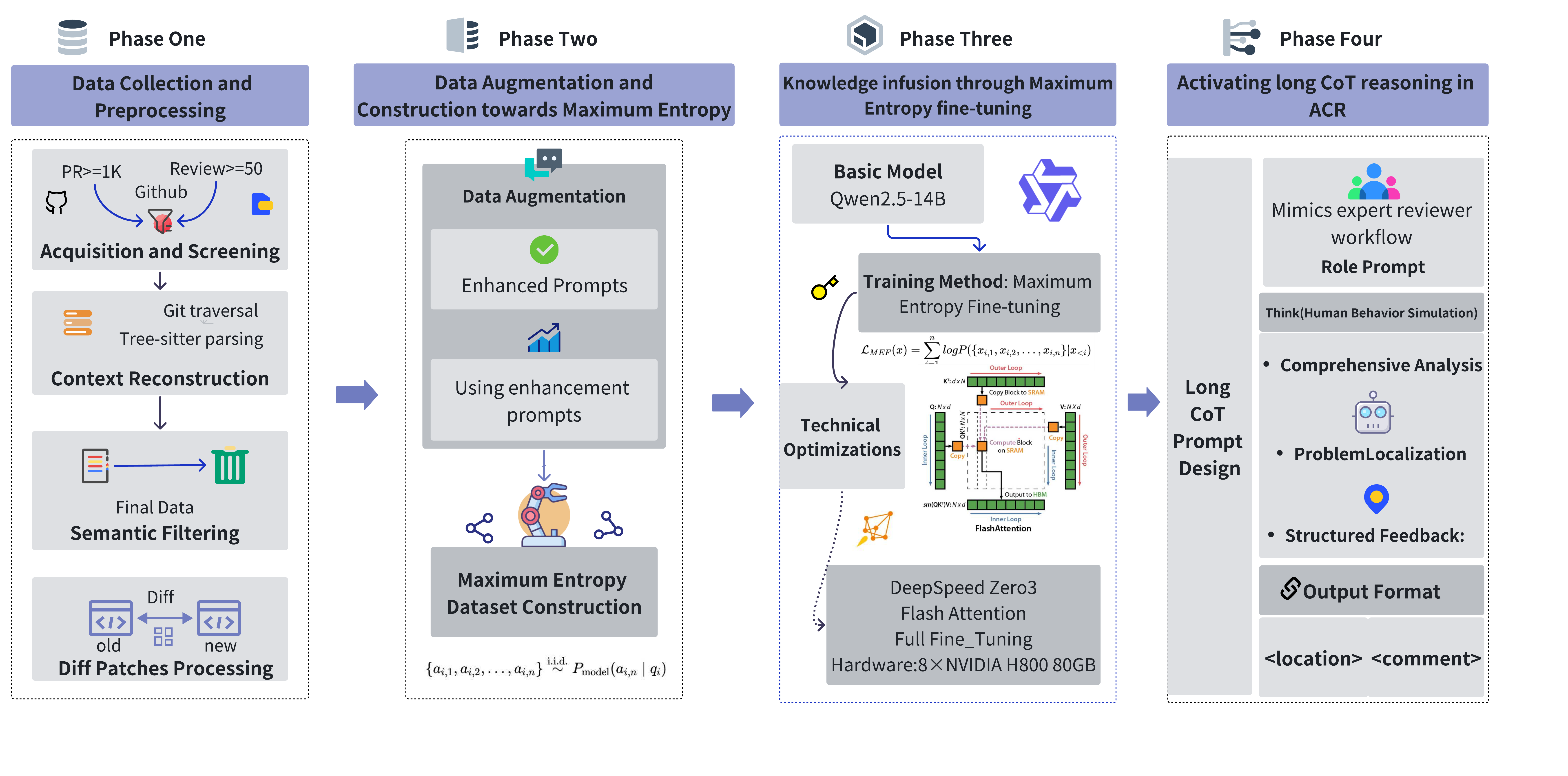}
    \caption{A schematic overview of the \MCot\ approach}
    \label{fig:overview}
\end{figure*}

Long COT extends this idea to more complex, multi-step tasks in which the model is asked to engage in deeper reasoning, considering multiple stages or layers of a problem before reaching a conclusion~\cite{chen2025towards, xia2024beyond, wang2025multimodal}. Edward Yeo et al.~\cite{yeo2502demystifying} explored the mechanics of long COT reasoning in LLMs and found that long COT enhanced reasoning capabilities by enabling strategies such as backtracking and error correction. The researchers systematically investigated the conditions under which long COT emerged, highlighting the role of reinforcement learning (RL) in developing these capabilities. Not only in the field of coding, but also in many other non-coding fields such as translation, long COT has demonstrated extremely strong capabilities. Jiaan Wang et al.~\cite{wang2024drt} introduced DRT, a new method that used long COT reasoning to enhance neural machine translation (MT). Ruohong Zhang et al.~\cite{zhang2024improve} introduced a COT reasoning approach to improve visual-language models, which significantly enhanced performance in visual tasks by augmenting training data and incorporating reinforcement learning. It has been found that a longer reasoning process enables the model to handle tasks with more intricate dependencies, ultimately providing better solutions.

\subsection{Maximum Entropy Methods}
Traditional RL typically aims to maximize cumulative reward~\cite{kaelbling1996reinforcement}. However, pursuing only high rewards can lead to overly ``deterministic'' policies, sacrificing exploration capability~\cite{arulkumaran2017deep}. To address this issue, researchers began to incorporate the idea of ME into RL~\cite{haarnoja2017reinforcement, schulman2017equivalence, haarnoja2018soft}. Soft Actor-Critic (SAC), proposed by Haarnoja et al.~\cite{haarnoja2018soft}, is currently one of the most representative ME-based RL algorithms. It is an off-policy deep RL algorithm that maximizes both the expected cumulative reward and the entropy of the policy. 

The principle of Maximum Entropy (ME) originated in information theory~\cite{jaynes1957information}. Its core idea advocates that, given known constraints, the probability distribution with the greatest entropy (informational uncertainty) should be chosen to avoid inherent bias and local optima. In short, ME fundamentally shifts the objective of policy optimization in RL to ``maximize reward + maximize entropy''~\cite{levine2018reinforcement}.
This strategy shows impressive efficacy in RL, e.g., in text classification and machine translation, ME classifiers avoid overfitting and effectively integrate diverse features by maximizing the entropy of the conditional probability distribution~\cite{nigam1999using, el2015arabic, och2002discriminative, ittycheriah2005maximum}. These early applications have laid the groundwork for extending the ME principle to complex decision-making systems.

A long COT prompt can also be viewed as a continuous and complex decision-making process. Furthermore, during an ACR process—as previously discussed—there exists an inherent need to simultaneously consider multidimensional information and perform comprehensive analysis. These characteristics demonstrate substantial comparability with features already observed in current ME-based solutions.

\section{Methodology}
\label{sec:meth}
Figure~\ref{fig:overview} depicts the \MCot\ approach, which comprises four phases: (1) collection and filtration of code review data from the open source community, (2) construction of ME-regulated fine-tuning datasets, (3) knowledge infusion through ME-regulated fine-tuning, and (4) activation of long COT reasoning capabilities through custom-crafted prompts.

\subsection{Data Collection and Preprocessing}
This phase includes the main steps such as raw data acquisition and initial screening, historical context reconstruction, semantic filtering, and long code `diff' truncation.

\paragraph{Step 1: Raw data acquisition and screening} We collected project development data from the GitHub Archive\footnote{https://www.gharchive.org/}, which captures timeline data of open-source projects including code submissions, review requests, and merge operations. Our data retrieval focused on code review records spanning from January 1, 2022 to November 1, 2024, with explicit filtering of review comments that triggered actual code modifications later on. This filtering strategy aligns with the established research consensus~\cite{rong2024distilling, kononenko2016code, widyasari2023explaining} that reviews leading to substantive fixes tend to contain valuable information for code review. As a result, the initial dataset comprised 12,152,191 raw review comments. To select active projects that can provide more abundant information for the dataset, we established the selection criteria that require a minimum of 1000 pull requests and 50 review comments. This selection process yielded 11,324 qualified projects containing 6,735,961 filtered review comments.

\begin{figure}[htbp!]
    \centering
    \includegraphics[width=0.9\textwidth]{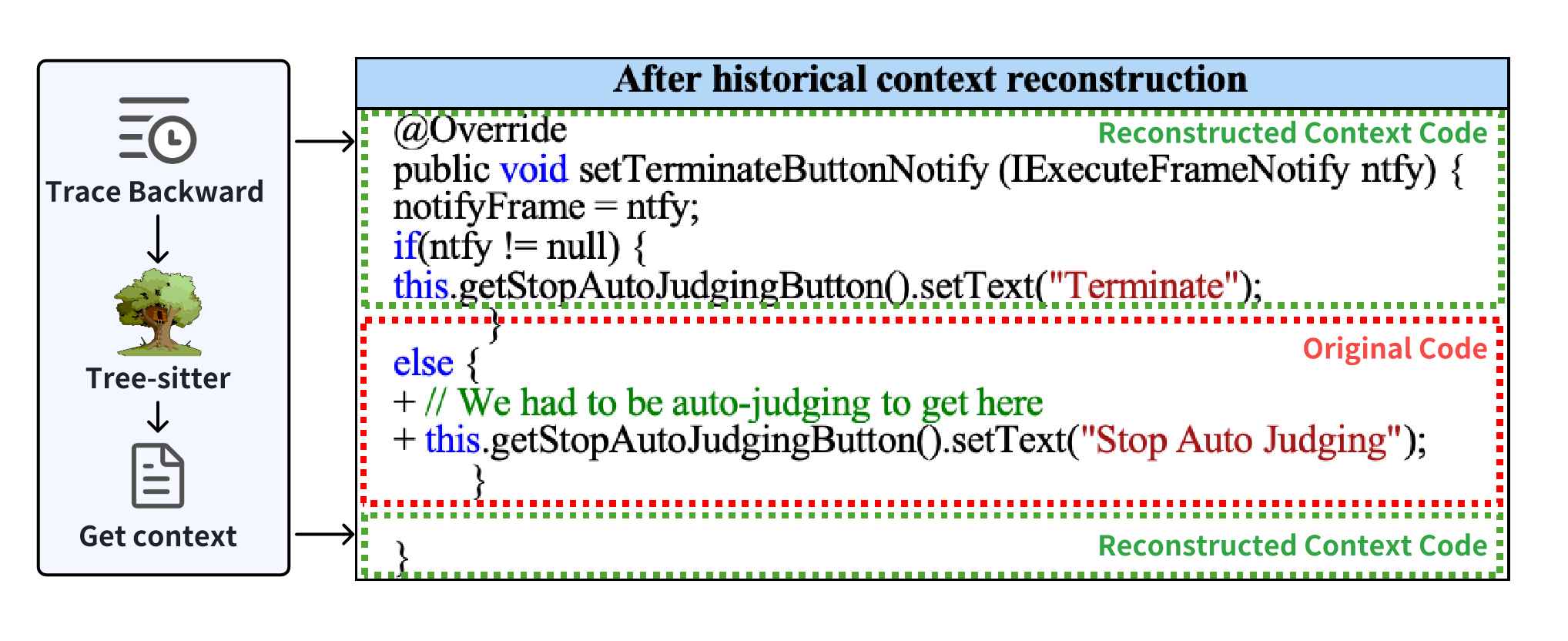}
    \caption{The historical context reconstruction}
    \label{fig:extractExample}
\end{figure}

\paragraph{Step 2: Historical context reconstruction} As illustrated in Figure~\ref{fig:extractExample}, we reconstructed the historical project states by traversing version-controlled files, as reviewers typically inspect code within its necessary development context. A critical challenge arises from the inherent limitations of using isolated `diff' patch fragments referenced in review comments in current studies~\cite{yu2024fine,lu2023llama}: a single `diff' fragment typically contains only incomplete partial code logic. However, we observed that when feeding isolated `diff' fragments to LLMs for code review, LLMs tend to disproportionately focus review attention on the `diff' fragment, identifying issues located in or related to that `diff' segment. Yet, code issues do not always reside within `diff' fragments. 

To address this validity concern, we cloned 11,324 open-source repositories and systematically reconstructed comprehensive `diffs' through git commit analysis. Our method consists of two substeps. First, we traced backward through commit histories associated with each review, matching the review's referenced `diff' fragments against historical commits to identify the precise pre-commit state. Subsequently, we employed tree-sitter~\footnote{https://github.com/tree-sitter/tree-sitter} to establish semantic context for each `diff': for intra-function modifications, we extracted the encompassing function body as context; for extra-function changes, we captured the syntactic unit containing the `diff' (e.g., class definitions, struct declarations, or module-level scopes). It should be noted that our parsing process specifically targeted mainstream programming languages (Java, Python, C/C++, JavaScript/TypeScript, and Go)~\cite{yu2024fine} to ensure practical relevance, resulting in 211,868 valid data entries. Notable data attrition occurred during this process, primarily due to two factors. The first was that historical commit truncation may occur to manage repository size, preventing recovery of early development information. The second was our systematic exclusion of documentation-related commits during code parsing, despite their prevalence in open-source reviews. 

\paragraph{Step 3: Semantic filtering}

Previous studies~\cite{yang2023evacrc, rong2024distilling, yu2024distilling, chen2025understanding} have noted the presence of low-value reviews in raw peer review comments, which also occurred in our dataset. To improve data quality, we needed to filter out these review comments. Using similar strategies applied in a previous study~\cite{yu2024fine}, we performed semantic filtering on our dataset, resulting in 13,881 data entries. Table~\ref{tab:comment} lists typical types of low-value comments.

\paragraph{Step 4: Long code truncation}

To control resource consumption during training, we performed truncation on excessively long code while strictly preserving complete `diff' fragments. Using the QWen2.5 tokenizer~\cite{qwen2025qwen25technicalreport}, we calculated tokens and enforced a 1000-token limit. When the code exceeded 1000 tokens, we retained three lines of code preceding and three lines following the `diff' fragment as its context, maintaining the integrity of the `diff' itself. After this step, we obtained a curated dataset of 12,881 code review instances. We refer to this dataset as the \MCot\ dataset.

\begin{table}[h!]
    \centering
    \scriptsize
     \caption{Typical low-value comments}
    \label{tab:comment}
    \begin{tabular}{|c|l|}
    \hline
        \makecell[c]{Categories of \\Low-Value Reviews} & Descriptions  \\
        \hline
        Confirmation & \makecell[l]{Developer acknowledgments of issue resolution, \\predominantly containing status updates rather than \\technical insights}  \\
        \hline
        Submission Notices & \makecell[l]{Automated bot-generated messages referencing\\ commit hashes, filtered through hash value patterns\\ via regular expressions} \\
        \hline
        Pull Request Events & \makecell[l]{Notifications regarding PR merges or \\initiations devoid of technical problem descriptions} \\
        \hline
        URL References & \makecell[l]{Comments containing unreachable web links or \\external resource pointers excluded due to\\ crawling limitations}  \\
        \hline
        Mentions & \makecell[l]{@-based user notifications lacking technical problem \\statements or contextual details} \\
        \hline
        Test Suggestions & \makecell[l]{Recommendations for test additions without \\sufficient contextual justification, generally \\considered non-critical} \\
        \hline
    \end{tabular}
\end{table}

\subsection{Data Entry Augmentation and Construction}

There are two major steps in this phase: (1) data entry augmentation, and (2) maximum entropy dataset construction.
\paragraph{Step 1: Data entry augmentation}

\begin{figure}[h!]
    \centering
    \includegraphics[width=5.0in]{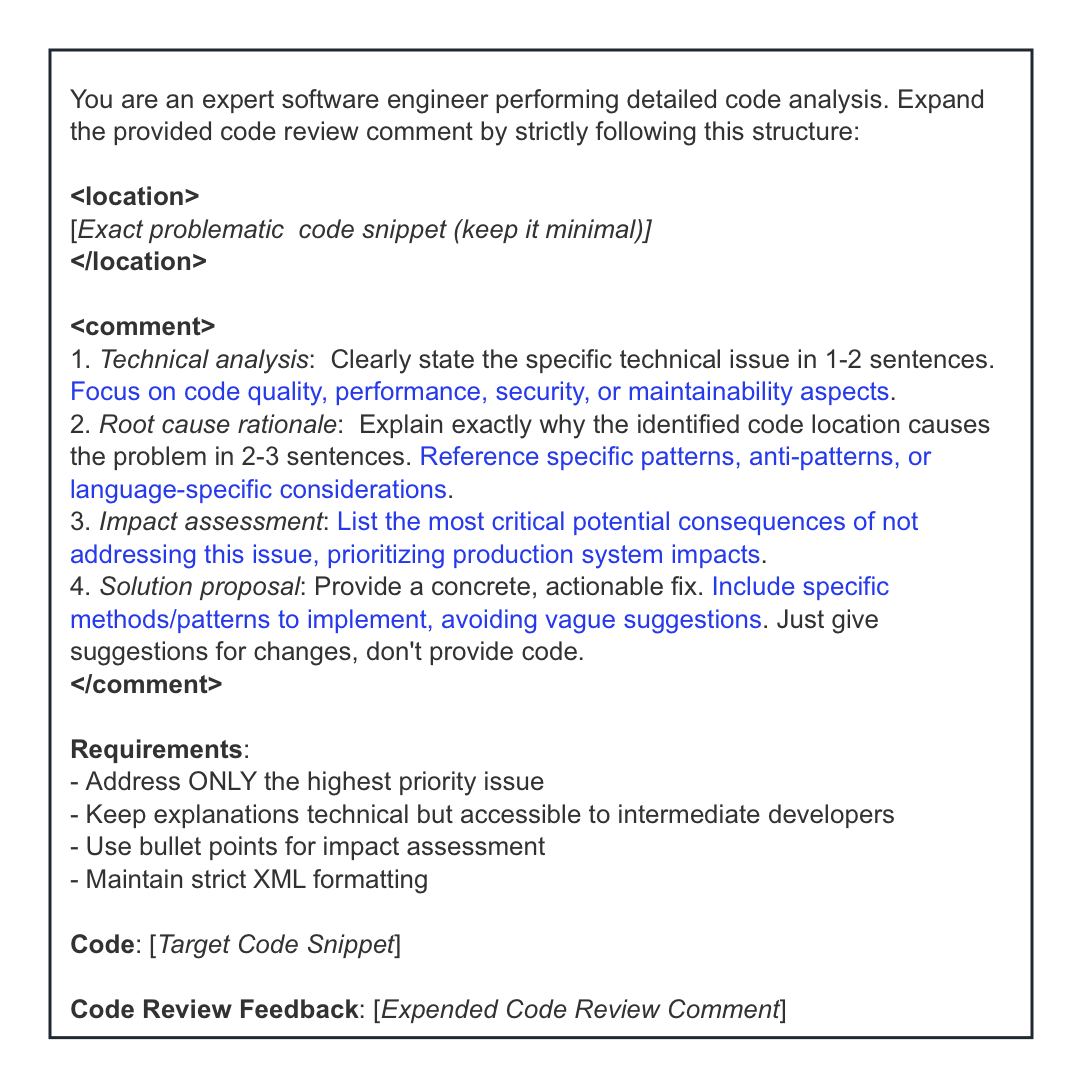}
    \caption{Prompt used to generate ME-regulated fine-tuning datasets}
    \label{fig:EnhancePrompt}
\end{figure}

Original comments retrieved from open-source projects mostly describe issues directly, but do not articulate the full analytical reasoning process that leads to the conclusion of those issues. This phenomenon is detrimental to model training as valuable information is missing, a limitation confirmed by previous research~\cite{yu2024fine}. So we also performed logical augmentation on the original review comments. Specifically, we needed to 
enhance both the expression and logic of the original review comments, which should include clear problem locations, professional explanations and root cause analyses of the issues, an assessment of the potential impacts of the issues, and the suggested solutions. The specific enhancement prompts are shown in Figure~\ref{fig:EnhancePrompt}.

%\vspace{-.5cm}

\paragraph{Step 2: ME-regulated fine-tuning dataset construction}
A conventional approach to constructing a fine-tuning dataset, as formulated in Equation~\ref{eq:conNormal}, involves pairing each query \( q \) with a canonical answer \( a \), thereby guiding the model to generate responses aligned with predefined standards during subsequent fine-tuning processes. 
\begin{equation}
    a_i \sim P_{\text{model}}(a_i \mid q_i)
\label{eq:conNormal}
\end{equation}
However, this conventional fine-tuning method leads the models to mimicking both the desired correct answers and their solution paths only from the training data. 
Although this limitation is less problematic in conventional COT, it may exert negative effects in long COT scenarios because the latter, as the term suggests, inherently involves multiple coherent reasoning steps demanding logical self-consistency.
To address this limitation, we emphasize infusing essential knowledge about code review into the model while eliminating trivial elements such as expression formats or styles. To ensure content validity while maximizing response diversity, we formulate a ME-regulated fine-tuning data construction paradigm shown in Equation~\ref{eq:conME}. For each query \( q \), we generate \( n \) (set to 10 in our study) distinct answer instances \(\{a_1, a_2, ..., a_n\}\) to comprehensively model the ME probability space. 

\begin{equation}
     \{a_{i,1}, a_{i,2}, \ldots, a_{i,n}\} \stackrel{\text{i.i.d.}}{\sim} P_{\text{model}}(a_{i,n} \mid q_i)
\label{eq:conME}
\end{equation}
The data entry generation process used the QWQ 32B model~\cite{qwq32b} as the base model. All inference parameters strictly adhered to QWQ 32B's recommended configurations~\cite{qwq32b} to ensure generation quality and diversity. Following the answer generation phase, we proceeded to construct instruction tuning data specifically designed for ME-regulated fine-tuning. The instruction template, as illustrated in Figure \ref{fig:EnhancePrompt}, was populated with code commit information and raw code review data to formulate a complete instruction, which was later used to generate enhanced code review comments.

\subsection{Knowledge infusion through ME-regulated fine-tuning (MEFT)}
This section provides MEFT details, including base model selection, loss function definition, and hyperparameter configuration.

\paragraph{Base model selection}Given the requirements of long COT reasoning, which necessitate extended context lengths and robust foundational model capabilities, we selected Qwen2.5-14B~\cite{qwen2025qwen25technicalreport} - one of the most powerful open-source models under 20B parameters in the open-source community - as our base model. This choice was constrained by computational resource limitations that precluded training larger models.

\paragraph{MEFT loss function definition} During the fine-tuning phase, we employed full-parameter fine-tuning to maximize the model's knowledge capacity for comprehensive review expertise integration. Conventional fine-tuning methods typically optimize the following objective:

\begin{equation}
     \mathcal{L}_{LLM}(x)= \sum^n_{i=1}log P(x_i|x_{<i})
\label{eq:finetune}
\end{equation}
while $P(x_i|x_{<i})$ means the model predicts the i-th token conditioned on preceding tokens. While this approach enhances the model's ability to generate standard answers, it risks overfitting through incidental learning of task-irrelevant priors present in reference answers, thereby constraining adaptability to complex tasks. To address this limitation, we implemented MEFT using diversified answer distributions:
\begin{equation}
     \mathcal{L}_{MEFT}(x)= \sum^n_{i=1}log P(\{x_{i,1}, x_{i,2}, \ldots,x_{i,n}\}|x_{<i})
\label{eq:MEfinetune}
\end{equation}
In general, this method enables the model to learn multiple solution pathways per problem, enhancing its generalization capabilities.

\paragraph{Hyperparameter configuration}For computational efficiency, we applied Flash Attention~\cite{dao2022flashattention, dao2023flashattention2} in the attention layers to reduce memory overhead from extended token sequences. Further memory optimization was achieved through DeepSpeed Zero3~\cite{10.1145/3394486.3406703}, which offloads the optimizer's states and partial model parameters to system memory. The complete training process was conducted on 8×NVIDIA H800 80GB GPUs. Detailed hyperparameter configurations are provided in Table~\ref{tab:TrainingHyp}.

\begin{table}[htbp!]
\caption{Training hyperparameters}
\begin{center}
\scriptsize
\begin{tabular}{|c|c|c|c|c|}
\hline
epochs & batch & lr & cutoff & warmup \\
\hline
2 & 64 & 1e-7 & 5000 & 500 \\
\hline
\end{tabular}
\end{center}
\label{tab:TrainingHyp}
\end{table}

\subsection{Activating long COT reasoning in ACR}
\begin{figure}[h!]
    \centering
    \includegraphics[width=5.0in]{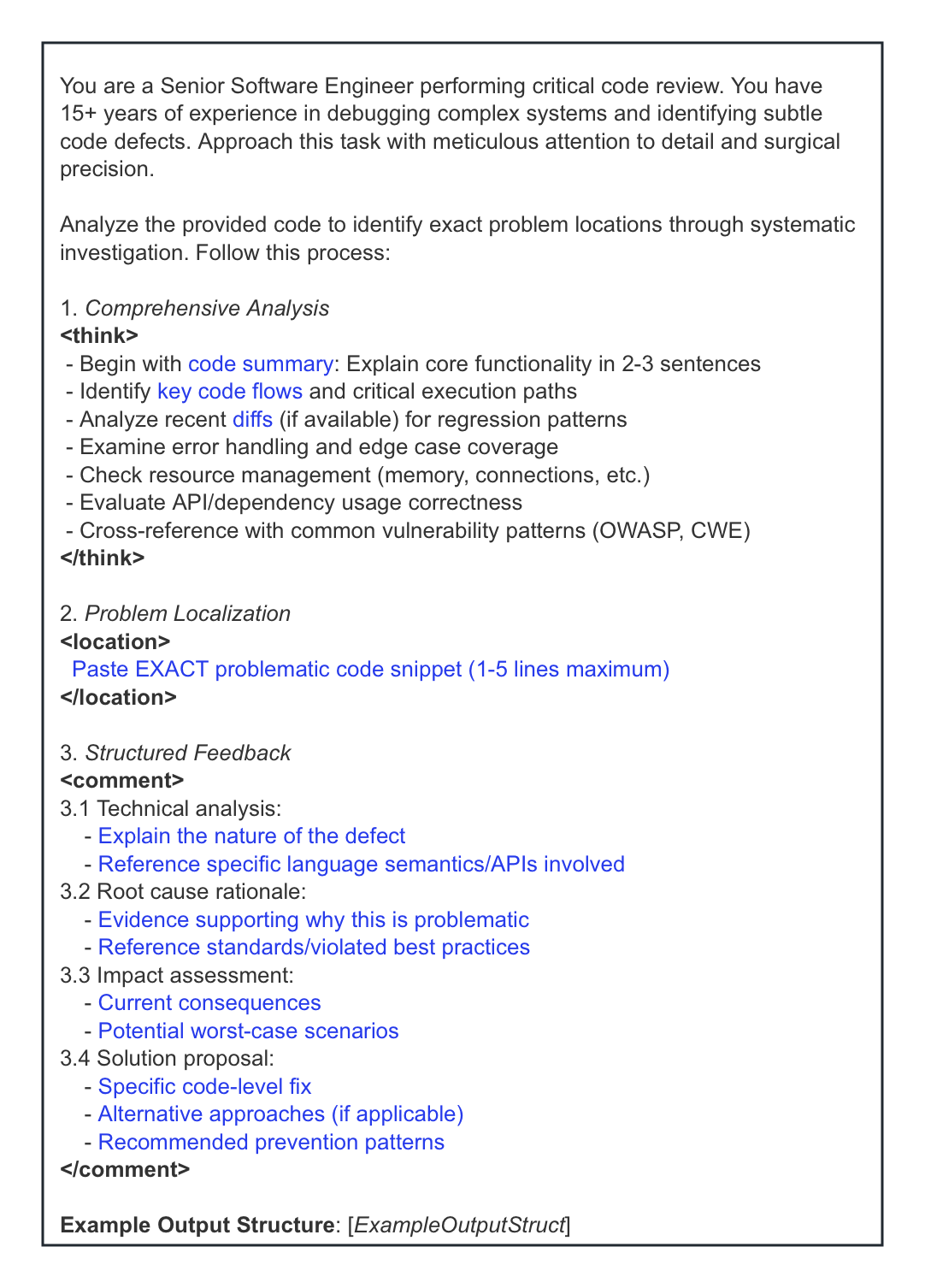}
    \caption{Prompt used to activate long COT reasoning in ACR}
    \label{fig:LCotPrompt}
\end{figure}

Existing research suggests that long COT improves the performance of LLMs by enabling them to engage in long reasoning processes and explore a broader range of considerations~\cite{chen2025towards, xia2024beyond, wang2025multimodal}. In the context of ACR, long COT mirrors a typical cognitive workflow of a human reviewer. First, they need to understand the implemented functionality and primary execution paths of the code under review. Then, they systematically examine the modifications introduced in the new submission to assess whether and how the changes impact the intended functionality or outcomes. After that, they conduct thorough analyses of potential code quality issues such as examining error (or exception) handling and edge case coverage, checking resource management (memory, connections, etc.), evaluating the correctness of API/dependency usage, and inspecting common vulnerability patterns~\cite{kononenko2016code}. So we constructed a long COT code review prompt to emulate this manual process, as illustrated in Figure~\ref{fig:LCotPrompt}. 

% \vspace{-.5cm}

\section{Evaluation}
\label{sec:eval}
To empirically validate the proposed \MCot\ approach, we formulate the following research questions:
\begin{itemize}
    \item RQ1: How effective is an LLM fine-tuned with \MCot\ in performing ACR tasks compared to existing methods?
    \begin{itemize}
        \item RQ1.1: How accurately does the fine-tuned model localize code issues?
        \item RQ1.2: How accurately do the review comments generated by the fine-tuned model describe the identified code issues?
    \end{itemize}
    \item RQ2: What factors critically impact \MCot's effectiveness in fine-tuning LLMs for ACR tasks?
\end{itemize}

In particular, RQ1 aims to validate that fine-tuning an LLM with a unique reasoning ability to analyze multiple dimensions of code review by \MCot\ can lead to improved ACR performance in terms of its accuracy in both localizing code issues and describing the identified issues. RQ2 aims to provide a deep understanding of the factors in \MCot\ that play critical roles in improving ACR performance.

In this section, we first describe the experimental settings and then analyze the results to answer the research questions.

\subsection{Experimental settings}
Experimental settings include datasets, benchmark models, and evaluation metrics.

\subsubsection{Datasets}The evaluation dataset is divided into two parts to evaluate the model performance in in-distribution and out-of-distribution review tasks, respectively. The in-distribution dataset consists of 1,000 randomly sampled test items from our curated \MCot\ dataset (the remaining data allocated to model training), assessing performance under the same distribution as the training data. The out-of-distribution dataset utilizes the CodeReviewer dataset~\cite{li2022codereviewer}. As the collection time periods of CodeReviewer and \MCot\  datasets do not overlap, it is a reasonable benchmark dataset to evaluate the out-of-distribution performance of \MCot. As the CodeReviewer dataset lacks issue location annotations, it is only included in assessing the accuracy of review comments.

\subsubsection{Benchmark models/approaches}
Benchmark models/approaches are drawn from two primary sources. One category comprises models fine-tuned specifically for ACR tasks. To our knowledge, the only publicly known studies are Carllm~\cite{yu2024fine}, LLaMA-Reviewer~\cite{lu2023llama}, BitsAI-CR~\cite{sun2025bitsai}, and CodeReviewer~\cite{li2022codereviewer}. However, we exclude CodeReviewer~\cite{li2022codereviewer} and LLaMA-Reviewer~\cite{lu2023llama} from our comparative analysis due to two fundamental limitations. First, their training protocol solely employs raw review comments without incorporating code location annotation capabilities. Second, Yu et al.'s work~\cite{yu2024fine} revealed their substantial performance gaps compared to current state-of-the-art benchmarks. Meanwhile, although BitsAI-CR~\cite{sun2025bitsai} is also designed to improve LLMs' ACR performance, it primarily concentrates on the data flywheel in enterprise environments. In addition, it utilizes internal corporate data and rules, making it impossible for us to conduct comparisons without a publicly available replication package.

The other category involves general models that are not fine-tuned for ACR tasks. Due to access restrictions, we cannot compare with advanced models like the GPT family. Instead, we selected accessible alternatives with comparable capability: Qwen 2.5 72B, QWQ 32B, and DeepSeek R1 671B. The benchmark models/methods are listed in Table~\ref{tab:model_information}.

\begin{table}[h!]
    \centering
    \scriptsize
\caption{Models/Approaches in RQs}
\label{tab:model_information}
    \begin{tabular}{|c|c|c|c|}\hline
         Models/Approaches&  Type&  Parameters& Research Questions\\\hline
 \textbf{\MCot}& Fine-tuned model& 14B&RQ1, RQ2\\\hline
         Carllm&  Fine-tuned model&  14B& RQ1\\\hline
         Qwen 2.5 72B&  General model&  72B& RQ1\\\hline
         QWQ 32B&  General model&  32B& RQ1, RQ2\\\hline
          DeepSeek R1 (0120)&  General model&  671B& RQ1\\ \hline
    \end{tabular}
\end{table}
\vspace{-.5cm}

\subsubsection{Objective Metrics}
We use the following objective metrics to evaluate the performance of various ACR approaches.

\uline{\emph{Intersection over Union (IoU):}}
To evaluate code issue localization performance, we employ the \emph{Intersection over Union (IoU)} metric~\cite{rahman2016optimizing, rezatofighi2019generalized}, which is calculated as follows:
$$
IoU=\frac{label \cap \mbox{\it predict}}{label\cup \mbox{\it predict}}
$$
where $\mbox{\it label}$ denotes the set of code lines containing issues in the ground truth, and $\mbox{\it predict}$ represents the set of code lines identified as problematic in the prediction. A higher \emph{IoU} value signifies more precise issue localization, with the metric reaching its maximum value of 1 when the predicted and the ground-truth code segments completely overlap. This measurement effectively quantifies the spatial alignment between actual and detected code issues, providing an objective criterion for evaluating issue localization accuracy.

\uline{\emph{Hit Rate:}}
This evaluation metric is calculated as follows:
$$ Hit~Rate = \frac{hit ~count}{total} $$
where \emph{hit count} denotes the number of instances where the model-generated review comments align with the issues identified in the ground truth, and \emph{total} represents the full test set size of 1,000 entries.

\subsubsection{Human Evaluation Metrics}
The human evaluation process involves a comprehensive examination of three elements for each entry: the original code submission, historical review comments, and LLM-generated feedback. This tripartite analysis aims to determine the practical value of the generated reviews. The assessment framework comprises two distinct tasks:
\begin{itemize}
    \item \textit{Accurate Identification of Historical Reviews}: Requires the LLM-generated comments to describe identical issues to those specified in the ground truth. 
    \item \textit{Provision of Valuable Review Feedback}: Mandates that the identified issues must objectively exist in the code samples without containing hallucinatory content.
\end{itemize}

Human evaluation was conducted by two senior students specializing in software engineering. From the complete test set, 540 entries were randomly sampled for manual assessment, ensuring a 95\% confidence level with less than 3\% margin of error. Each evaluator analyzed 300 entries, including 60 overlapping samples for consistency verification. The inter-rater reliability analysis yielded a Cohen's kappa~\cite{kvaalseth1989note} score of 0.8426, indicating substantial agreement. We also examined the consistency between LLM-as-judge~\cite{li2024llms, zheng2023judging, gu2024survey} and human evaluation. The calculated kappa value was 0.5281, also indicating substantial agreement between human evaluation and LLM-as-judge evaluation. 

To facilitate statistical measurement, analysis, and comparison, we define the following two metrics:

\underline{\emph{Human Hit:}} involves manually comparing the review comments generated by the model with the standard answers to determine whether the code issues identified by the model align with the ground-truth answers. The final result is calculated as the proportion of consistent cases out of the total number of evaluations.  
$$ Human ~Hit = \frac{human~hit ~count}{total} $$

\underline{\emph{Human Valuable:}} involves manually reviewing the model-generated comments alongside the original code to assess whether the model has identified valuable code issues. Here, ``valuable" is defined as the model pointing out genuine code problems, regardless of whether they match the ground-truth answers. The final result is calculated as the proportion of valuable review comments out of the total number of evaluations. %\R{to add equation here}
$$ Human~ Valuable = \frac{human~valuable~count}{total} $$

\begin{figure}[h!]
    \centering
    \includegraphics[width=5in]{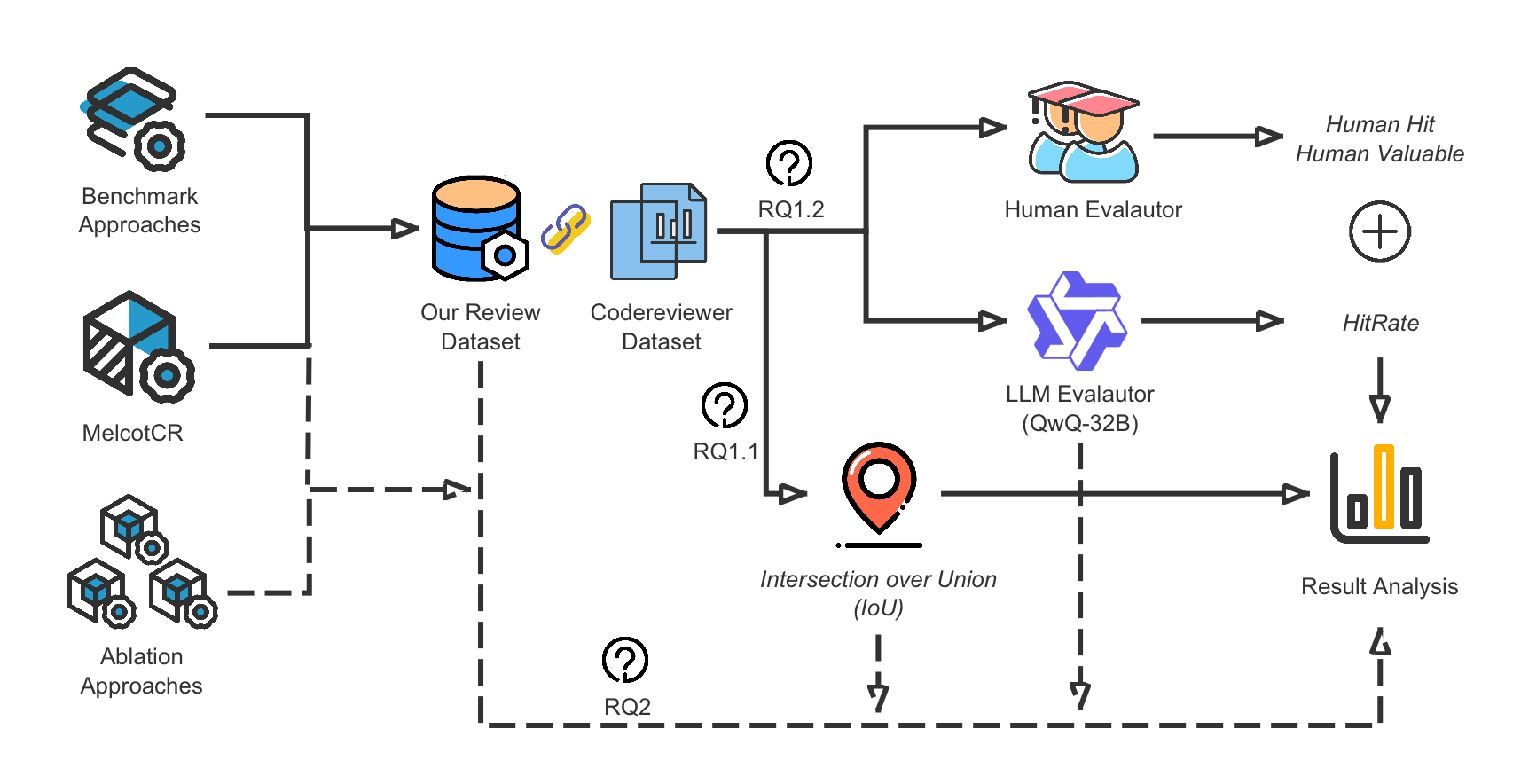}
    \caption{The process for comparative evaluation of \MCot}
    \label{fig:RQProcess}
\end{figure}

\subsubsection{The experimental evaluation process}
Figure~\ref{fig:RQProcess} depicts our comparative framework consisting of two primary components. For the first category concerning prior code review studies, we select the state-of-the-art baseline Carllm~\cite{yu2024fine} as our benchmark. To ensure methodological parity, we replicate the experimental setup by maintaining identical model parameter scales, equivalent training data scope, and consistent implementation of full fine-tuning during model reproduction. The second comparison focuses on contemporary dominant LLMs utilizing prompt engineering for review generation. We choose QWQ-32B~\cite{qwq32b}, Deepseek-R1 671B~\cite{guo2025deepseek}, and Qwen2.5 72B~\cite{qwen2025qwen25technicalreport} as baseline references. During the evaluation process, we conduct iterative prompt optimisation to achieve optimal generation quality, with explicit instructions requiring models to prioritize the identification of critical issues to prevent low-quality review outputs.

\begin{table}[htbp!]
    \centering
    \scriptsize
\caption{Experimental settings for RQs}
\label{tab:experiment_setting}
    \begin{tabular}{|c|c|c|c|c|}\hline
         RQ&  Datasets&  Methods&  \makecell{Evaluation\\ methods}& Metrics\\\hline
         1.1&  \MCot&  \makecell{Carllm 14B  \\ \MCot\ 14B \\ Qwen2.5 72B \\ QWQ 32B \\ Deepseek-R1 671B}&  automated& IoU\\\hline
         1.2& \makecell{ \MCot\ and\\ CodeReviewer }&  \makecell{Carllm 14B  \\ \MCot\ 14B \\ Qwen2.5 72B \\ QWQ 32B \\ Deepseek-R1 671B}& \makecell{ Human and \\automated }& \makecell{Hit Rate, Human Hit and \\Human Valuable}\\\hline
         2&  \MCot&  \MCot\ 14B&  automated& IoU and HitRate\\\hline
    \end{tabular}
\end{table}

\underline{The experiment for RQ1.1}
Localizing code issues plays a crucial role in ACR as it helps developers pinpoint the exact locations of code issues, 
and improves both reviewing and fixing efficiencies. As shown in Table~\ref{tab:experiment_setting}, the evaluation of code issue localization employs \emph{IoU} for automated assessment. This metric provides a detailed evaluation of the distance between the positions of code issues identified by the model and the ground truth.

\begin{figure}[h!]
    \centering
    \includegraphics[width=5in]{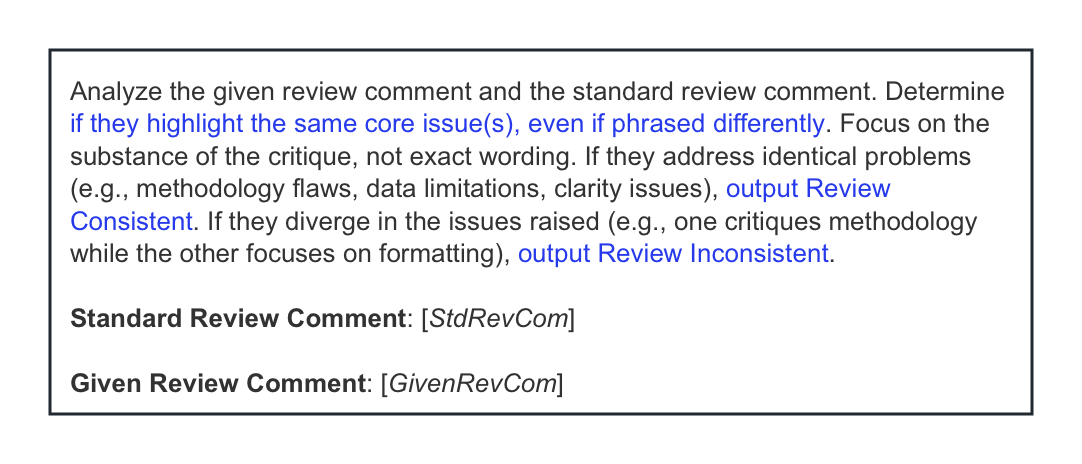}
    \caption{LLM-as-Judge prompt }
    \label{fig:evaluatePrompt}
\end{figure}

\underline{The experiment for RQ1.2}
Code review comments exhibit diverse expression methods, which leads to significant bias when using similarity metrics such as BLEU~\cite{papineni2002bleu} to evaluate the performance of review comments, as many studies have reported~\cite{tran2019does, kocmi2024navigating}. As shown in Table~\ref{tab:experiment_setting}, to avoid evaluation bias caused by expression differences, we employ a combination of LLM-as-Judge~\cite{li2024llms, zheng2023judging, gu2024survey}, commonly used for assessing complex semantic tasks, and manual evaluation to measure the accuracy of generated review comments. For LLM-as-Judge, we assess whether the generated review comments identify the same issues as the reference answers, using the prompt shown in Figure~\ref{fig:evaluatePrompt}. When constructing the evaluation prompt, we compare the generated comments with the reference answers to determine the correctness of the review. By providing the correct answers, we guide the LLM to focus on the content of review comments rather than their expression, thereby significantly mitigating subjective bias in LLM evaluations.  
% \vspace{-.5cm}

The evaluation dataset has two parts: 1,000  samples randomly selected from the \MCot\ dataset to evaluate performance on in-distribution data and 1,000 entries randomly selected from the CodeReviewer dataset~\cite{li2022codereviewer}. During data sampling, we select code languages consistent with our study to avoid potential bias from less common languages. As the CodeReviewer dataset cannot provide ground truth for code issue locations, we only conduct manual and automated evaluations of the accuracy of generated review comments when using this dataset. The automated evaluation employs the QWQ 32B model as the base model, which demonstrates performance comparable to Deepseek-R1 while being more computationally efficient. 

\underline{The experiment for RQ2} 
The process of analyzing which factors in \MCot\ significantly impact its performance in fine-tuning LLMs for ACR tasks comprises two main parts. The first part examines the effect of ME. The second part explores the effect of the thought process. As shown in Table~\ref{tab:experiment_setting}, we conduct ablation experiments on different factors through automated evaluation, focusing on two aspects: positional localization and accuracy of code review. When analyzing the impact of ME on ACR performance, we compare ME-regulated fine-tuning against conventional fine-tuning with the options of regular and long COT arrangements. To assess the influence of different thinking steps on the final ACR performance, we investigate the impact of each individual step on the overall task by removing specific step prompts.

\subsection{Evaluation results}
Following the evaluation settings and process elaborated in the previous section, this section analyzes the evaluation results. 

% \vspace{-.2cm}
\begin{table}[htbp!]
    \centering
      \caption{Performance of each method on the \MCot\ and CodeReviewer datasets}
    \label{tab:performance}  
    \scriptsize
    \begin{tabular}{|c|c|c|c|c|c|c|c|}
    \hline
     & \multicolumn{4}{c|}{\MCot\ dataset } & \multicolumn{3}{c|}{CodeReviewer dataset} \\
    \hline
         &  IoU&  \makecell{Hit\\ Rate}&\makecell{Human\\Hit}& \makecell{Human \\Valuable}&  \makecell{Hit\\ Rate}&  \makecell{Human\\Hit}& \makecell{Human \\Valuable}\\
         \hline
         Carllm 14B  &  24.12&  21.9&  25.83& 76.11&  35.02&  36.40& 87.78 \\
         \hline
         \makecell{\MCot\ \\ (Qwen2.5 14B)} &  \textbf{27.16}&  \textbf{25.4}&  \textbf{29.72}& 81.67&  36.56&  39.17& \textbf{92.00}\\
         \hline
         Qwen2.5 72B&  10.26&  18.8&  15.56 & 73.05&  27.62&  29.17 & 75.28\\
         \hline
         QWQ 32B&  11.46&  22.7&  20.28& 78.33&  25.94&  25.28& 71.39\\
         \hline
         \makecell{Deepseek-R1 \\671B (0120)}&  12.17&  25.2&  26.11& \textbf{83.61}&  \textbf{38.38}&  \textbf{40.56}& 86.11\\
         \hline
    \end{tabular}
\end{table}

\subsubsection{RQ1.1: Issue localization accuracy} Table~\ref{tab:performance} presents the \emph{IoU} performance of different methods in positional localization on the \MCot\ and CodeReviewer datasets. The results confirm that the low-parameter Qwen2.5 14B base model, when fine-tuned with \MCot, significantly surpasses Carllm, the state-of-the-art model fine-tuned for ACR tasks so far, and all the other models, especially Qwen2.5 72B, the same series but with significantly more parameters and Deepseek-R1 671B, enabling developers to precisely identify problematic code segments. LLMs not specifically trained for ACR generally exhibit suboptimal performance. Manual investigation reveals distinct failure patterns between QWQ 32B and Deepseek-R1: QWQ 32B's lower score primarily stems from its inability to detect valid code issues, whereas Deepseek-R1 tends to over-expand localization scope, unnecessarily increasing code inspection workload for reviewers and consequently wasting their cognitive efforts.

% \R{Any insightful finding?}

\begin{figure*}[]
    \centering
    \includegraphics[width=\linewidth]{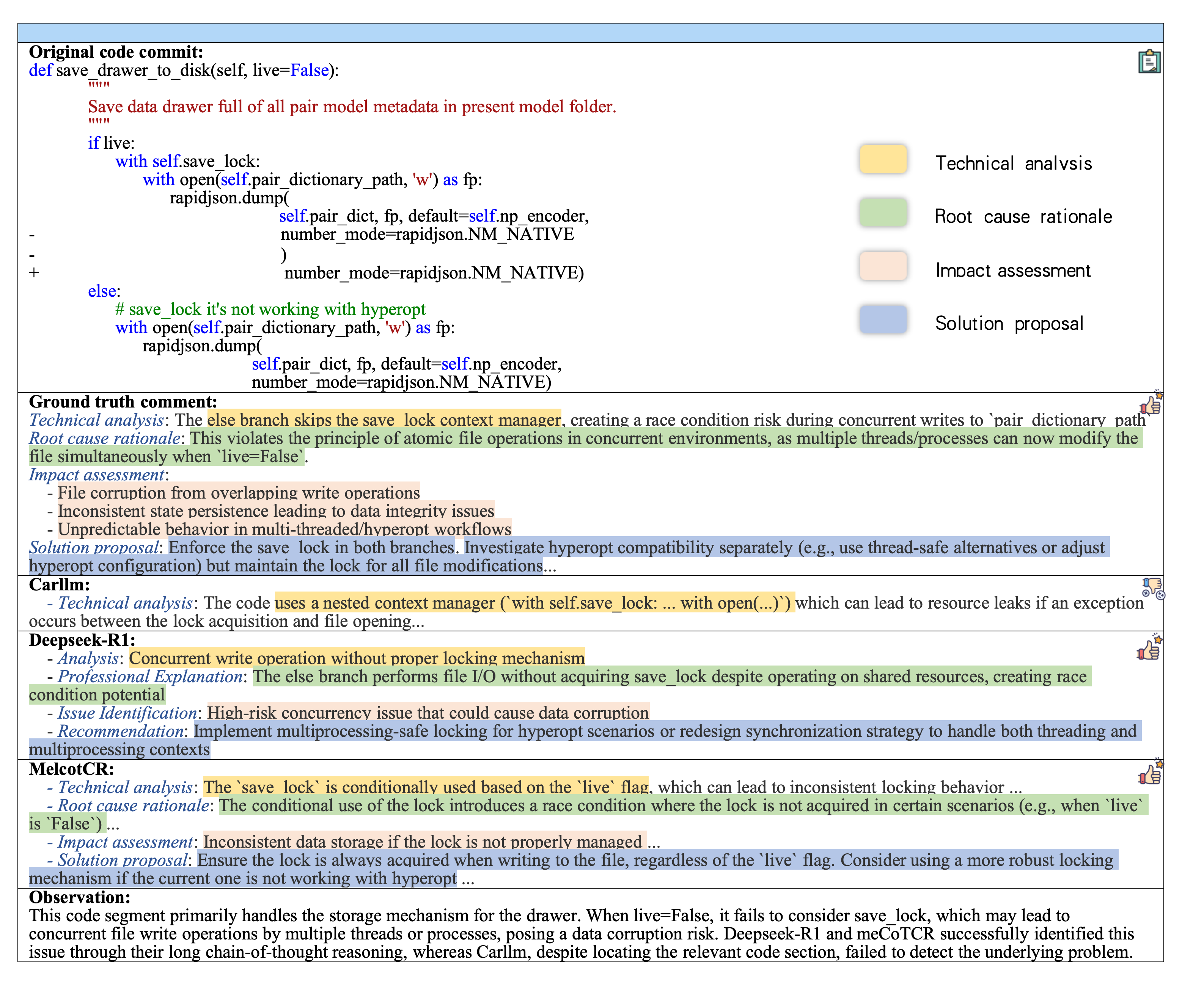}
    \caption{The example comments generated by Carllm, \MCot, and Deepseek-R1}
    \label{fig:LCotExample}
\end{figure*}

\subsubsection{RQ1.2: Issue description accuracy} Table~\ref{tab:performance} presents the results of both automated and human evaluations. For automated evaluation in terms of \emph{Hit Rate}, \MCot\ outperforms all other methods, including Deepseek-R1 whose parameter size is an order of magnitude larger on the \MCot\ dataset. This advantage stems from the comprehensive knowledge infusion through MEFT and the analytical steps designed to emulate the human review process. As illustrated by a comparative example in Figure \ref{fig:LCotExample}, \MCot\ exhibits stronger analytical depth and logical reasoning capabilities compared to Carllm using a conventional COT approach.

For human evaluation, \MCot\ outperforms all other methods, including Deepseek-R1, in accurately identifying code issues (in terms of \emph{Human Hit}), which confirms its effectiveness. In terms of maintaining the factual grounding of review comments to ensure that identified issues truly exist in code snippets without hallucination (\emph{Human Valuable}), \MCot\ surpasses all other methods including the state-of-the-art fine-tuned Carllm, confirming that combining ME-regulated fine-tuning with long COT reasoning leads to substantial improvement in code review capability. However, due to the significant parameter scale disparity between the base model and Deepseek-R1, \MCot\ slightly underperforms Deepseek-R1. 

When comparing LLM-as-judge with human evaluation results, QwQ 32B exhibits significant evaluation bias when assessing its own responses, displaying a higher probability of considering its own answers correct. Our human and LLM-based evaluations serve as mutual cross-validation to more accurately reflect the performance of different methods in generating reviews. Comparing different approaches reveals that the long COT model demonstrates superior performance in the review task. This further underscores the necessity of applying long COT techniques to ACR.

On the CodeReviewer dataset, \MCot\ performs comparably to Deepseek R1 on out-of-distribution data while significantly outperforming other methods. However, \MCot\ surpasses Deepseek R1 in generating useful review comments (\emph{Human Valuable}).

% \R{Any insightful finding?}

\subsubsection{RQ2: impacting factors} 
We examine factors influencing the performance of \MCot\ from two aspects of fine-tuning: fine-tuning methods and fine-tuning process steps.

\begin{table}[htbp!]
    \centering
        \caption{Effects of different fine-tuning methods and inferencing methods in \MCot}
    \label{tab:LCotE}
        \scriptsize
    \begin{tabular}{|c|c|c|c|}
\hline
Fine-tuning Method            & Inferencing Method & IoU   & Hit Rate \\ \hline
\multirow{2}{*}{Conventional} & Regular COT        & 24.12 & 20.0    \\ \cline{2-4} 
                              & Long COT           & 26.88 & 22.9    \\ \hline
\multirow{2}{*}{MEFT}         & Regular COT        & 26.60 & 20.3    \\ \cline{2-4} 
                              & Long COT           & 27.16 & 24.3    \\ \hline
\end{tabular}
\end{table}

\underline{\emph{The influence of fine-tuning methods.}}
As shown in Table~\ref{tab:LCotE}, MEFT exhibits distinct effects in localizing code issues and generating review comments. For the task of localizing code issues, MEFT with regular COT can achieve the same performance as the conventional fine-tuning that adopts long COT. When MEFT adopts long COT, its performance is elevated significantly. Even with the conventional fine-tuning method, adopting long COT makes a significant difference. For the task of generating precise review comments, MEFT does not yield significant improvement over conventional fine-tuning when using regular COT. In contrast, when employing long COT, MEFT demonstrates a notable advantage. As long COT incorporates additional outputs such as code analysis and change descriptions, it suggests that MEFT holds greater potential in enabling models to acquire deeper knowledge.

\underline{\emph{The influence of reasoning steps.}}
As shown in Table~\ref{tab:LcotStep}, the experimental results indicate that `diff analyze' has the greatest overall impact on correctly identifying code issue locations and generating accurate review comments, which aligns with the intuition that most code issues are triggered by code modifications~\cite{sadowski2018modern, bosu2014identifying}. `Summary' has the least influence, probably because summarization tasks are relatively simple, and high-level functional descriptions lack the granularity required to establish concrete connections with specific code defects, thus failing to provide sufficient contextual clues to improve accuracy in review tasks. `Key code flows' significantly affect the correct generation of review comments, suggesting that effective code review generation requires not only localized inspection but also a comprehensive understanding of how components interact within the broader context.
% \vspace{-.2cm}
\begin{table}[htbp!]
    \centering
\caption{Effects of different reasoning steps in \MCot}
\scriptsize
\label{tab:LcotStep}
    \begin{tabular}{|c|c|c|c|c|}\hline
         Steps&    IoU&Ratio&Hit Rate& Ratio\\\hline
         Full&    27.16&-&24.3& -\\\hline
         - Summary&    26.94&-0.81\%&23.9& -1.64\%\\\hline
         - Key code flows&    26.67&-1.80\%&22.3& -8.23\%\\\hline
         - Diff anlyze&    25.38&-6.55\%&22.7& -6.58\%\\\hline
         - Issue check&    26.32&-3.09\%&23.1& -4.93\%\\ \hline
    \end{tabular}
\end{table}
% \vspace{-.5cm}

% \R{Any insightful finding?}

\section{Discussion}
\label{sec:disc}
In this section, we discuss this work's implications for researchers and practitioners as well as its validity risks.

\subsection{Methodological significance of \MCot }
This study establishes three pivotal research directions for LLM-based ACR.

First, MEFT alleviates an issue in traditional supervised learning, that is, various biases are inadvertently introduced while acquiring knowledge. MEFT provides a new perspective: a shift from solely pursuing the best answer fitting to learning unbiased knowledge behind the answers.

Second, the long COT architecture establishes an extensible analytical framework for code review tasks. Unlike Yu et al.'s constrained three-phase decomposition of ``localization-description-resolution"~\cite{yu2024fine}, \MCot\ enables dynamic integration of new analytical dimensions including code summarization, core logic analysis, diff change analysis, and potential issue identification. This innovation not only expands the cognitive depth of LLMs in code review scenarios but also reveals a new research paradigm: enhancing code review performance through systematic integration of diverse perspectives to both understand and analyze source code.

Finally, this work empirically demonstrates that algorithmic innovations can enable smaller-scale models, e.g., 14B parameters, to match or surpass the performance of significantly larger counterparts, e.g., 671B-parameter Deepseek-R1. This discovery illuminates a promising direction for further research -- rather than continuously scaling up models, focusing on cognitive architecture enhancements could unlock the latent potential of existing models, thereby providing theoretical foundations for developing lightweight ACR systems.

% \vspace{-.3cm}
\subsection{Implications for practiontioners}
The superior performance of \MCot\ for ACR tasks demonstrates that incorporating multiple dimensions of information to facilitate the identification of issues in code by LLMs can yield positive gains. Through ME-regulated fine-tuning, we stimulate the model's long COT capability to effectively integrate multiple dimensions of information. This not only paves the way for the future evolution of ACR models but also leaves ample exploration space, for example, investigating whether introducing more dimensions of information or expanding the number of paraphrased review comments (this study used 10 semantically equivalent variants) could further enhance the model's deep reasoning capacity. We invite interested researchers to reproduce our work and explore this direction by making our replication package available at \url{https://anonymous.4open.science/r/MelcotCR}. 

It is also worth noting that the reasoning process of \MCot\ is inspired by the human review process and hence the \MCot\ approach is inherently friendly to human-AI collaboration in such a way that the tool's COT reasoning process provides auxiliary support for human reviewers to rapidly comprehend code functionality, core logic, critical modifications, and so on. This structured analytical approach assists reviewers in efficiently parsing code implementations, thereby creating additional efficiency gains in the code review workflow.

\subsection{Threats to Validity}
We are conscious that the following factors may introduce risks to the validity of the study's findings.

\underline{Limited LLMs.} Due to computational resource constraints, we experimented only with representative LLMs under restricted parameter configurations (less than 20B), specifically Qwen2.5-14B. Given the rapid evolution of LLM research, the broader community continues to explore the performance of diverse LLMs. Consequently, newer or larger models may outperform those evaluated in this work. In fact, existing studies have confirmed that long COT yields positive effects in certain scenarios only when the base model possesses sufficient parameters~\cite{li2025small}. However, the timing and environmental factors of model training limited our initial exploration to Qwen2.5-14B. Future work could investigate newer iterations such as Qwen3. We are open to experimenting with newer models.

\underline{Noisy Data.} Noise in the dataset may arise from multiple sources. First, while low-quality review comments in open-source communities were largely filtered out through semantic analysis, residual low-value data could still persist, potentially affecting model training. To mitigate this, we implemented manual screening and rule-based filtering to eliminate invalid reviews. Empirical results demonstrate that our approach maintains robust performance, suggesting minimal impact from residual noise.

\underline{LLM Evaluation Bias.} Prior studies have identified inherent biases in LLM-based evaluation, particularly in scoring tasks where models tend to favor content stylistically similar to their own outputs~\cite{li2025preference}. In this work, we adopt a contrastive approach to directly compare reference answers with model-generated outputs. By establishing explicit evaluation rules and objectives, our method significantly reduces model-specific biases.

\underline{Human Evaluation Bias.} Despite adopting standardized protocols and consistency checks, human evaluation can still introduce errors. To address this, we recruited evaluators with a software engineering background to ensure domain expertise in assessing the alignment between source code and review comments. We also conducted Cohen's kappa coefficient analysis to measure inter-rater reliability and permitted moderated discussions among evaluators to resolve contentious cases, further minimizing this risk.

\underline{Prompt Limitations.} While iterative experimentation was carried out to optimize prompt design, there may exist untested prompts capable of delivering superior performance. We encourage researchers to experiment with alternative prompts to explore further improvements.

\underline{Reasoning Steps.} It is generally believed that increasing COT length has a positive correlation with improved model performance. In this work, we enhanced COT reasoning length and achieved better performance through MEFT and explicitly specifying the model's reasoning path. It is worth validating whether increasing the reasoning length further can lead to even better performance.

\underline{10 Variations for Maximum Entropy.} Due to computational constraints, we provide only 10 variants for review comments, attempting to capture knowledge embedded in different pathways leading to the final answer -- correct review comments. However, the value 10 may not be necessarily optimal and it is worth exploring different values. We conjecture that increasing the number of variants would likely yield greater benefits, although an upper limit should exist to balance between the cost of resources and the performance gain. This awaits verification in subsequent research.

\section{Conclusions}
\label{sec:conc}
In this work, we propose \MCot, an approach that combines maximum-entropy-regulated fine-tuning and long chain-of-thought reasoning to optimize LLMs for ACR tasks by emulating the cognitive process of human code reviewers. Experimental results demonstrate that this method outperforms previous state-of-the-art approaches in both position localization and review comment generation in ACR tasks. With 14B parameters, it attains performance on par with Deepseek-R1 671B, one of the current strongest open-source long-chain reasoning models. Our future work includes leveraging online reinforcement learning to further enhance the review capabilities of LLMs, and developing context exploration methods for ACR systems to advance intelligent code review capabilities.

% \section*{Replication package}
%
% The replication package can be accessed via URL \url{https://anonymous.4open.science/r/MelcotCR}.

\bibliographystyle{ACM-Reference-Format}
\bibliography{Citations}
\end{document}